\documentclass[aps,preprint]{revtex4}
\def\BibTeX{{\rm B\kern-.05em{\sc i\kern-.025em b}\kern-.T
    08em\kern-.1667em\lower.7ex\hbox{E}\kern-.125emX}}
\usepackage{graphicx}
\usepackage{color}

\begin{document}
\title{Magnetic energy harvesting and concentration at distance by transformation optics}

\author{Carles Navau, Jordi Prat-Camps, and Alvaro Sanchez$^*$}
\affiliation{Grup d'Electromagnetisme, Departament de F\'{\i}sica, Universitat Aut\`onoma de Barcelona, 08193 Bellaterra,
Barcelona, Catalonia, Spain}


\begin{abstract}

Magnetic energy is one the main agents powering our society: generating energy in power plants, keeping information in magnetic memories, moving our devices with motors. All of these applications require a certain spatial distribution of magnetic energy, for example concentrating it in a transformer core or in a magnetic sensor. We introduce in this work a way to collect magnetic energy and distribute it in space with unprecedented efficiency and flexibility, allowing very large concentration of magnetic energy in a free space region, an enhanced magnetic coupling between two magnetic sources, and the transfer of magnetic energy from a source to a given distant point separated by empty space. All these features are achieved with a single device, a magnetic shell designed by transformation optics.

\end{abstract}

\maketitle

Transformation optics has pushed the possibilities of controlling light towards previously unexplored limits \cite{TOpendry,review_TO}, including perfect lenses \cite{perfectlens} and electromagnetic cloaks \cite{controlling,microwave}. When applied to static magnetic fields \cite{wood}, transformation optics ideas have allowed unique results such as the experimental realization of an exact cloak \cite{science}. An important application of transformation optics is concentration of electromagnetic energy, which is attempted using plasmonics \cite{schuller,aubry} or macroscopic concentrators \cite{rahm}. The former only concentrates at subwavelength scales (typically, nanometers for visible or infrared light \cite{plasmonics}) and the latter requires filling the concentration space with material \cite{rahm}. Here we apply transformation optics ideas to shape and concentrate magnetic fields in an unprecedented way. Differently from the electromagnetic waves case, our device will not require material in its interior so that it represents an ideal collector for energy harvesting or for increasing the sensitivity of a magnetic sensor placed inside.  The same device surrounding a magnetic source will enhance the field in the exterior. The combination of these two features allows to transfer the magnetic energy from a source to a desired distant point through free space. We should remark that this is a unusual effect because, different from the propagating electromagnetic fields, static magnetic fields naturally decay with the distance from the source (in the typical case of a magnetic dipole, as one over the third power of the distance). We will show that other unique properties can be achieved by combining our magnetic device with or without magnetic sources in their interior; for example, two magnetic sources each surrounded by our devices get an enhanced magnetic coupling, which may have applications in powerless transmission of energy.

. 

\section*{RESULTS}
\subsection*{Design of the magnetic shell}

Consider an infinitely long (along $z$-direction) cylindrical shell of interior and exterior radii $R_1$ and $R_2$, respectively, dividing the space in three domains:  interior ($\rho<R_1$), exterior ($\rho>R_2$) and shell ($R_1<\rho<R_2$) regions. To fulfill our goals, we will demonstrate that it is sufficient that (i) given a source of magnetic energy in the exterior, the shell concentrates all the magnetic energy that would be in the material transferring it to the hole, and (ii) if the source is in the interior, the shell energy is transferred to the exterior. 

We start by considering the magnetic source in the exterior domain; we want our shell to transfer all the shell energy into its hole. The required shell material can be determined by transformation optics technique \cite{TOpendry,review_TO}; linearly compressing the region from $\rho=0$ to $\rho=R_2-\xi$ in the interior domain and potentially expanding the annulus from $\rho=R_2-\xi$ to $\rho=R_2$ in the shell region (see Supplementary Information Sect. I). In the $\xi \rightarrow 0$ limit no space is left in the shell region which means that all the energy is concentrated inside. As the exterior space is not transformed, the exterior magnetic field (and energy) distribution is unaffected. 
This allows to understand why magnetic energy is totally transferred from the shell to the hole: exterior magnetic energy is unaltered whereas in the shell energy is zero (from the transformation it can be seen that {\bf B} is radial and {\bf H} angular so its product ${\bf B}\cdot{\bf H}$, the magnetic energy density, is zero), so it can only go to the hole. The material in the shell has to be homogeneous and anisotropic with radial and angular relative permeabilities $\mu_{\rho}\rightarrow \infty$ and $\mu_{\theta}\rightarrow0$.
Interestingly, despite the interior space is transformed, no material is needed in the hole (unlike the electromagnetic waves case \cite{rahm}). 
For such a shell in a uniform applied field, the field in the hole is uniform, has the direction of the applied field {\bf H$_0$}, and its magnitude is increased with respect to $H_0$ by a factor $\frac{R_2}{R_1}$,
\begin{equation}
\label{absolute}
	H^{HOL} = H_0 \frac{R_2}{R_1}.
\end{equation}
Similar results can be obtained for an actual 3D shape, like a spherical shell.
The magnetic behavior of the shell can be seen in Fig. 1a; a large field concentration is achieved inside while keeping the external field not distorted. 

One can arrive to Eq. (\ref{absolute}) by another path. 
From Maxwell equations, we have analytically solved the general case of a homogeneous and anisotropic shell in a uniform applied field (see Supplementary Information Sect. III). We remark two results. First, the cases with $\mu_{\theta}=1/\mu_{\rho}$ are those that do not distort the external field \cite{antimagnet}. Second, for a given $R_2/R_1$ ratio, the field in the hole increases when increasing $\mu_\rho$ (for fixed $\mu_\theta$) and also when decreasing $\mu_\theta$ (for fixed $\mu_\rho$). The absolute maximum for magnetic field concentration is achieved in the limit $\mu_\rho\rightarrow\infty$ and $\mu_\theta\rightarrow 0$, so it corresponds to a non-distortion case, and has the value given in Eq. (\ref{absolute}). 

\subsection*{On the practical realization of the magnetic shell}

Concerning the practical realization of our device, our concentrator requires $\mu_{\rho} \rightarrow \infty$ and $\mu_{\theta} \rightarrow 0$; actual materials with such anisotropy do not exist. But an approximation consisting of an alternation of radially displaced ferromagnetic and superconductor wedges (or even rectangular prisms) will constitute a natural discretization of the required material as the ferromagnets will give the large radial permeability and the alternated superconductors will cancel the angular components of {\bf B} field, leading to an effective $\mu_{\theta}=0$, as demonstrated in Figs. 1b-d. 
For a large $N$ -easily achieved in practice with thin sheets- the field in the hole is very homogeneous and approaches the exact limit. Such superconducting and ferromagnetic materials are readily available, even commercially. In \cite{science} we fabricated a magnetic cloak using these two materials, and their ideal behavior was experimentally confirmed for fields as large as 40mT and liquid nitrogen temperatures (in our present case the permeability of the magnetic layer should be larger, so permalloy or similar alloys could be used instead). 

For small applied field values as required for sensitive magnetic sensors, the behaviour of the real materials will be very close to the ideal case. For large applied fields, superconductors have been used to concentrate magnetic energy with standard procedures for applied fields up to several teslas in magnetic lenses \cite{lens}, so the applicability of our device for large fields will be mainly determined by the saturation field of the ferromagnets, at which its permeability may decrease towards smaller values than needed. 

\subsection*{Magnetic concentration and application to magnetic sensors}

Our results can be applied to increasing the sensitivity of magnetic sensors, like SQUIDs, magnetoresistance or Hall sensors. Magnetic concentration is typically used to enhance their sensitivity \cite{robbes,pannetier,ripka,lenz2006,griffith,SQUIDreview}, 
by either ferromagnetic materials, attracting magnetic flux, or diamagnetic ones, repelling it, being superconductors the optimum diamagnetic materials. The usual concentration strategy is based on two superconductors -or two ferromagnets- separated by a gap in which flux concentration is produced. We show in Figs. 2a-b a typical example of concentration of a uniform applied magnetic field in the gap of two superconducting slabs (assumed ideal, with zero permeability $\mu$ \cite{science}). The field gets enhanced at the edges of the slabs, not only on the gap region but also on the exterior ones. When reducing the slabs thickness, the contribution from the edges adds up and a rather intense magnetic field develops in the gap (Fig. 2b), although the gap field is not homogeneous but increases towards the strip edges and decreases at the central region. For our shell with permeabilities $\mu_\rho\rightarrow\infty$ and $\mu_\theta\rightarrow 0$, an applied magnetic field will be enhanced homogeneously in the shell hole by a factor $\frac{R_2}{R_1}$ (Eq. (\ref{absolute}), so a magnetic sensor placed there would detect a much larger field -increasing its sensitivity by a large known factor. 
By comparing with the superconductors case, in our design not only the average field in the hole is always larger (Fig. 2c), but also the magnetic flux is always constant whereas for the gap strategy the flux tends to zero (Supplementary Fig. S7) \cite{clem}, because field lines are mainly diverted toward the exterior edges when the gap is narrowed. 
Our concentration design is optimum because all the magnetic energy enclosed in the material region is totally transferred to the hole.
Another case of interest is for sensing non-uniform fields from nearby magnetic sources, as in biosensors \cite{biosensors}, in measuring human brain response in magnetoencephalography \cite{pizzella,fagaly} or for detecting single magnetic microbeads \cite{microbead}. Often in these situations it is measured not the field but its gradient, in order to separate the signal of the source from other distant noise sources. Transformation optics allows to analytically obtain
the magnetic field at any point of the hole and its gradient (Supplementary Information Sect. IB). The gradient becomes scaled by a higher power of the radius relation, [$(R_2/R_1)^{2}$ for the gradient of a dipolar source], which makes our concentrator particularly useful for magnetic gradiometers. 
 
\subsection*{Magnetic energy expulsion and concentration at distance}

We now study the case when the magnetic source is placed inside; all the magnetic energy in the shell has to be expelled towards the exterior domain. The required material is designed by transforming the space so that the shell space is totally pushed out. Interestingly, applying linear and potential transformations we find that the required shell is exactly the same homogeneous anisotropic shell with $\mu_{\rho}\rightarrow \infty$ and $\mu_{\theta}\rightarrow0$ that we designed for energy concentration above (Supplementary Information Sect. II).

The shell property of transferring all the magnetic energy in the shell to the exterior when there is a source in the hole is illustrated in Fig. 3. The magnetic energy of a magnetic dipole (e. g. a small magnet) in free space (Fig. 3a) is distributed to a further radial distance by enclosing the dipole with our shell (Fig. 3b). It is worth to notice that outside the shell the field is exactly the same that would be if there was a centered dipole in the empty space, with a magnetic moment enhanced by a factor $R_2/R_1$. In particular, this enhanced magnetic moment can ideally be made arbitrarily large by making $R_1$ arbitrarily small. Although being counterintuitive, this can be understood if we take into account that the shell expels the magnetic energy that would be in the shell volume towards the exterior, and a small $R_1$ implies that the shell extends to points near the dipole, where the energy density arbitrarily increases.

When adding a second shell at a given far distance, the enhanced field of the first dipole is harvested by the second shell and concentrated in its hole (Fig. 3c): magnetic energy has been transferred from the dipole to a desired position through empty space. Analytical expressions for the field in all these cases are obtained by the corresponding space transformations (Supplementary Information Eqs. S28 and S13).

\section*{DISCUSSION}

Many combinations of shells with or without inner sources may be constructed based on these ideas. As an example with possible interest for applications, we show in Fig. 4 how the magnetic coupling between two dipoles separated by a given free-space gap can be substantially enhanced by separating them and surrounding them with two of our shells. This enhanced magnetic coupling may have relevance to power wireless transmission \cite{soljacic}, where a key factor for achieving this goal is to increase the mutual inductive
coupling between the source and the receiver
resonators \cite{smith}. Although our ideas strictly apply to static fields, extension to other frequency ranges such as those used in power wireless transmission would be possible if materials with the required permeability values at such non-zero frequencies are found, possibly using metamaterials \cite{wiltshire}.

Another example of interest is shown in Fig. 5, where we present a strategy to increase the magnetic energy created by a collection of sources (four magnets in this example) in a region of free space. By surrounding each of the four magnets with our magnetic shells, a large concentration of magnetic energy is obtained in the central region (red colour in Fig. 5c), with a value much larger than not only that for the original 'naked' magnets (Fig. 5a) but also that for the same magnets when they are brought to a closer distance (Fig. 5b). These ideas could be applied to medical techniques like transcranial magnetic stimulation (TMS) \cite{trans}, for which large magnetic fields are required at a given position in the brain, which is not accessible for magnetic sources. As explained in \cite{bonmassar}, TMS generally targets superficial cortical regions, and deeper targets, such as the basal ganglia, are beyond the range of current technology; our ideas may help extend the reach of magnetic fields toward deeper depths in the body.

Finally, we comment on the main physical idea behind the results in this work. As mentioned in the introduction, devising new ways of distributing magnetic energy in space is key for developing new applications and improve existing ones. Superconductors and ferromagnetic materials have been traditionally used for shaping magnetic fields, separately. In terms of magnetic energy, both exclude the magnetic energy from their interior (ideal superconductors because {\bf B}=0 and ideal ferromagnets because {\bf H}=0, so that magnetic energy density ${\bf B}\cdot{\bf H}$ is zero in both cases). Our device, both in the ideal case of a homogeneous anisotropic material and in its discretized versions (Fig. 1), also expels the magnetic energy from its interior, because in this case {\bf B} is always perpendicular to {\bf H} in the material. The new property that we exploit in this work is that when the topology is such that the device divides the space into an exterior and an interior zone (e. g. a hollow cylinder or a sphere), then the energy of the magnetic field is totally transferred from one domain to the other (e. g. from an external source to the hole of the cylinder or sphere), contrary to the case of only superconductors and ferromagnets, where the energy is returned to the domain where the magnetic sources are (Supplementary Fig. S8). This is the key physical fact that has allowed the results of this work, and future configurations that may be developed based on these ideas as well.

\section*{METHODS}
\subsection*{Numerical calculations}
All numerical calculations have been done using the electromagnetics module (magnetostatics) of Comsol Multiphysics software.

\section*{Acknowledgements}

We thank Spanish Consolider Project NANOSELECT (CSD2007-00041) for financial support.
JPC acknowledges a FPU grant form Spanish Government (AP2010-2556).

\section*{Contributions}

All authors equally contributed to this work.

\section*{Competing financial interests}

The authors are inventors on a Universitat Autonoma de Barcelona patent application.

\section*{Corresponding author}

Alvaro Sanchez (alvar.sanchez@uab.cat)


\newpage

\begin{figure}[htbp]
	\centering
		\includegraphics[width=0.7\textwidth]{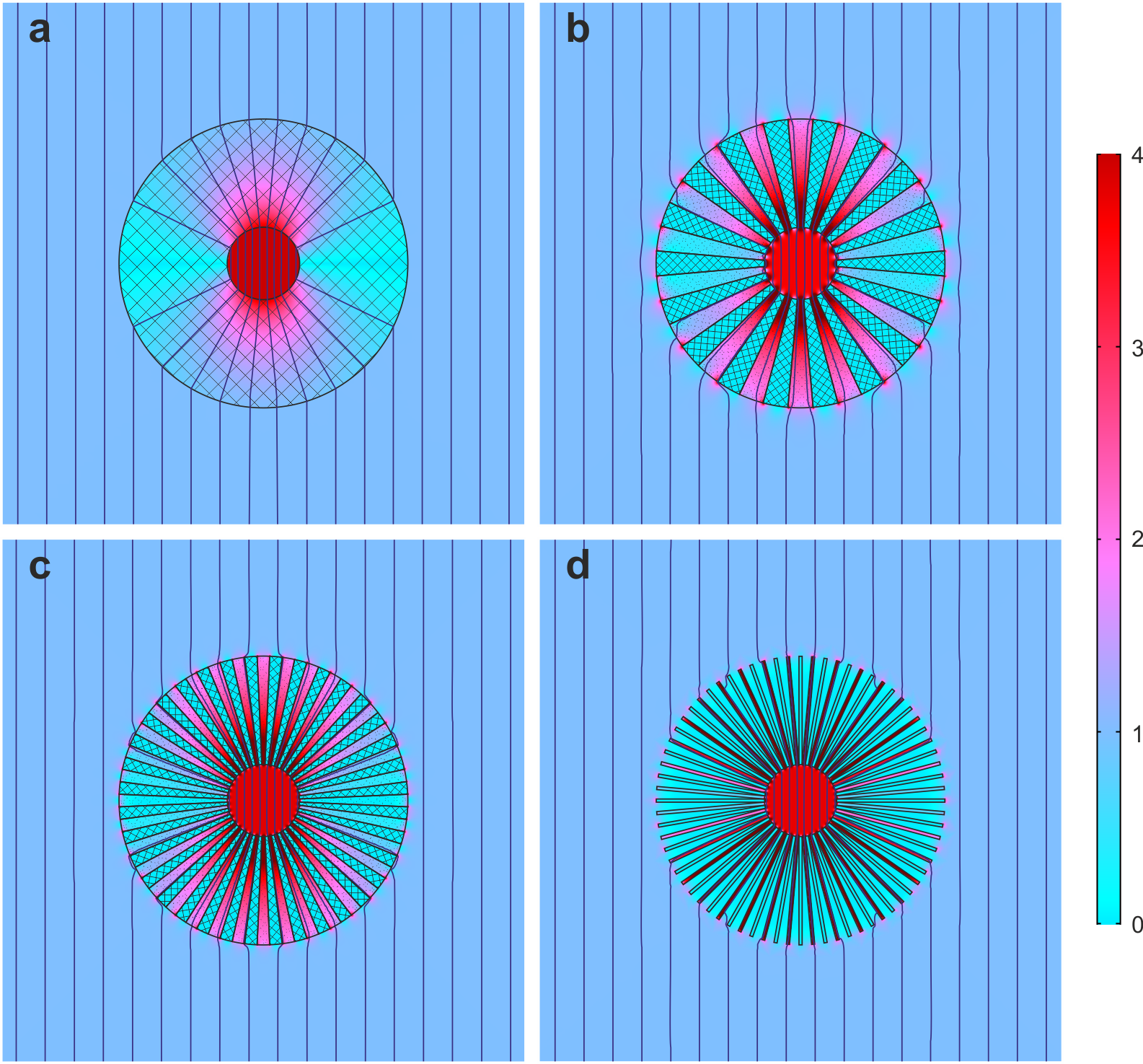}
	\caption{{\bf Ideal and discretized versions of the magnetic concentrator.} Magnetic field lines and their density (in colour; in units of applied magnetic field) for {\bf (a)} the ideal cylindrical magnetic concentrator shell with homogenous anisotropic permeabilities $\mu_{\rho}\rightarrow \infty$ and $\mu_{\theta}\rightarrow0$ and radii ratio $R_2/R_1=4$, and three discretized versions: {\bf (b)} 36 wedges of alternating homogenous and isotropic ideal superconducting ($\mu=0.0001$) and ideal soft ferromagnetic ($\mu=10000$) wedges; {\bf (c)} same with 72 wedges; and {\bf (d)} same with 72 rectangular prisms, for which a very good behavior is obtained even though materials do not fill the whole shell volume.  }
	\label{fig1}
\end{figure}

\begin{figure}[htbp]
	\centering
		\includegraphics[width=0.7\textwidth]{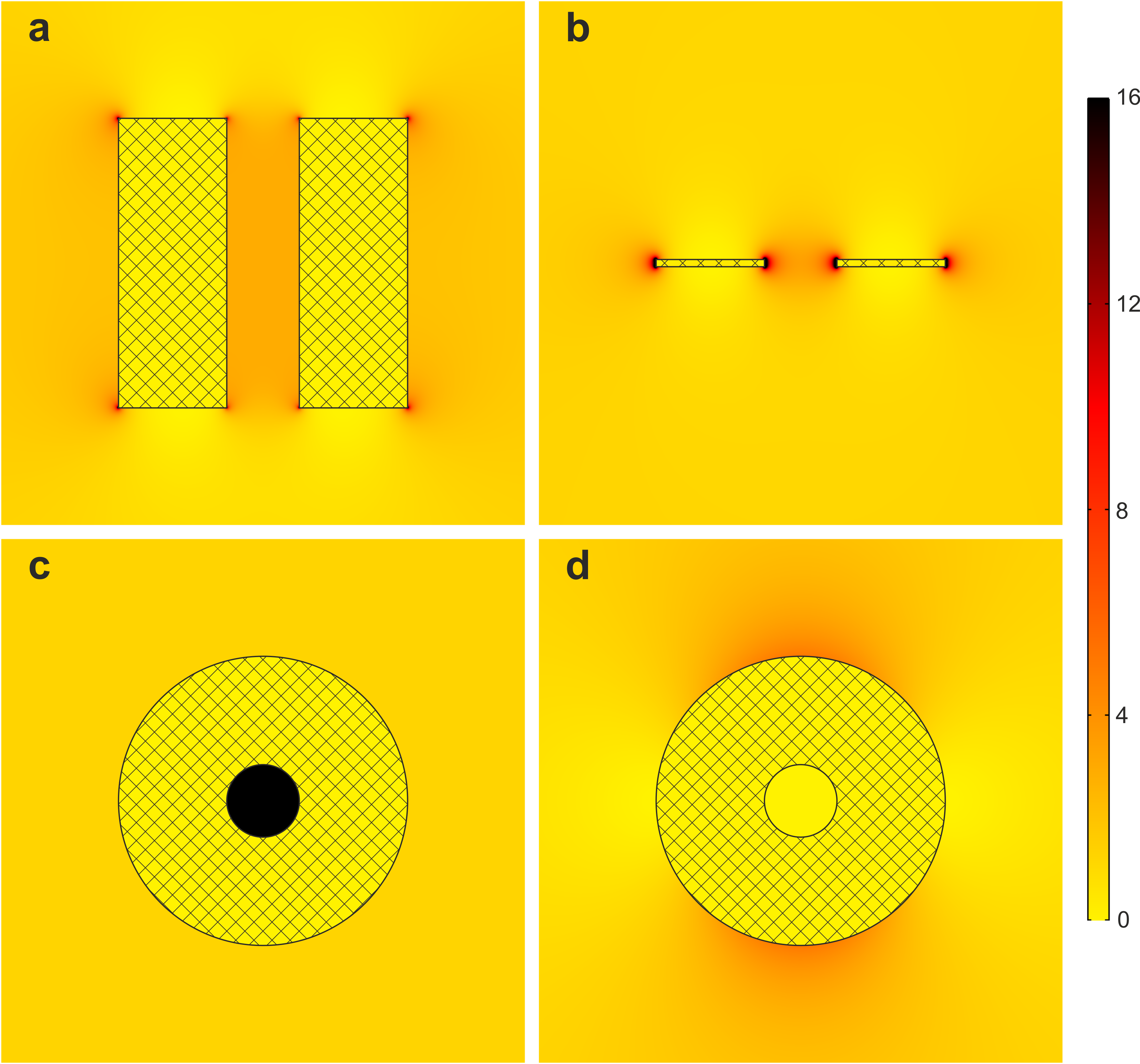}
	\caption{{\bf Optimum magnetic field concentration.} Common stategies for field concentration use the gap between two superconductors. In {\bf (a)} and {\bf (b)} we show the magnetic field energy density (normalized to that of applied field) for two ideal superconducting blocks ($\mu=0.0001$) with a thickness $w$ separated by a gap $b$ and decreasing heights $h$ = {\bf (a)} 4$b$, and {\bf (b)} $b$/10, in a vertical applied magnetic field. In {\bf (c)}, our optimum homogeneous anisotropic concentrating shell ($\mu_{\rho}\rightarrow \infty$ and $\mu_{\theta}\rightarrow 0$) with $R_2/R_1=4$ shows a large uniform energy density in its hole (for the comparison, notice that $2R_1=b$ and $w=R_2-R_1$).  In {\bf d}, a homogeneous isotropic shell made of ideal soft ferromagnetic material with the same dimensions as {\bf c} attracts more field lines from outside but cannot bring them to the hole, so the inner energy density is zero. } 
	\label{fig2}
\end{figure}

\begin{figure}[htbp]
	\centering
		\includegraphics[width=0.5\textwidth]{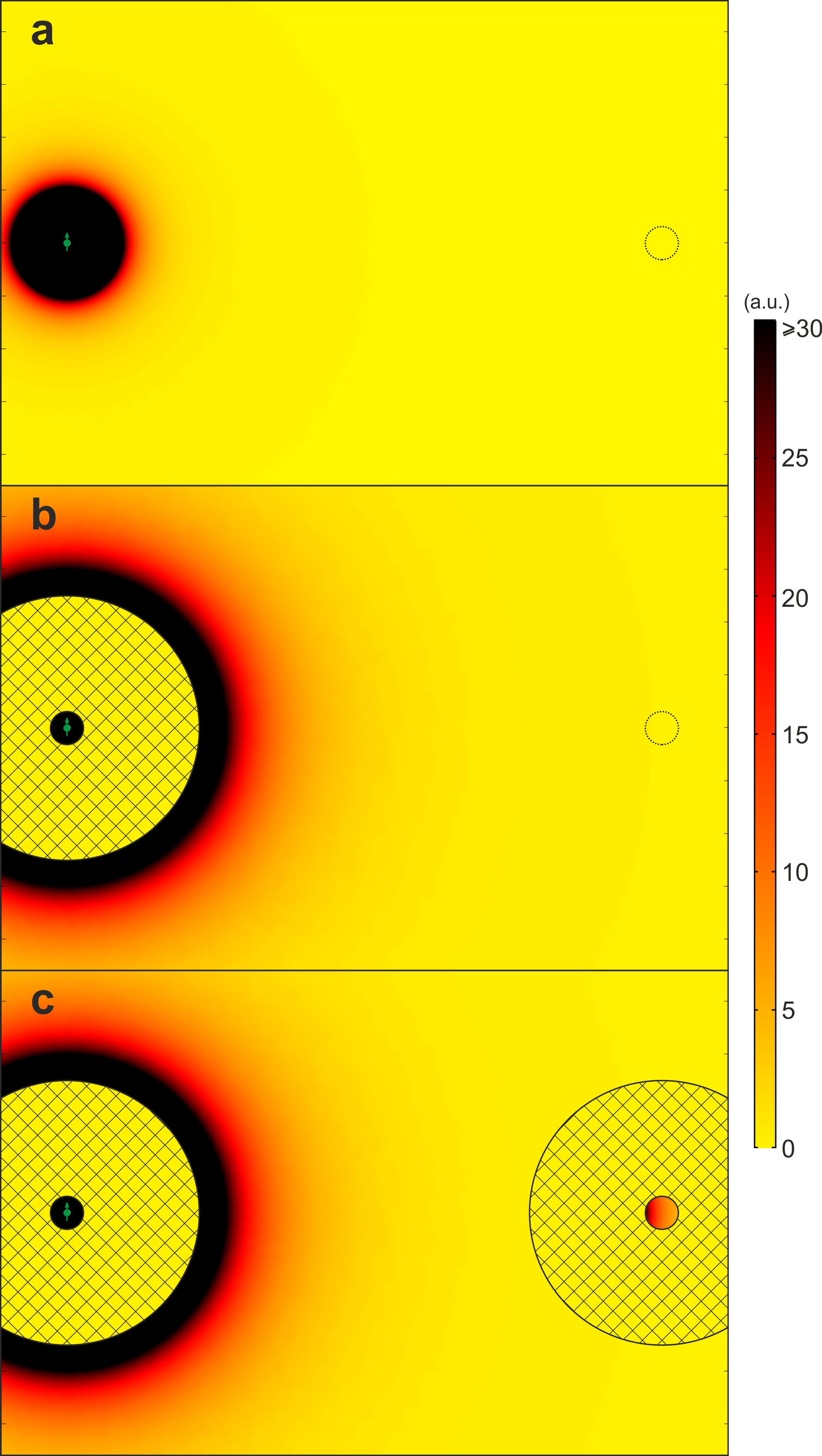}
	\caption{{\bf Magnetic concentration at distance through free space.} The magnetic energy density of a cilindrical dipole (e. g. a small magnet) {\bf (a)} is spatially redistributed towards larger radial distances by the use of our shell with $R_2/R_1=8$ {\bf (b)}. In {\bf (c)}, a second concentrator harvests the enhanced magnetic energy in its volume and transfers it to its hole, where a large value of magnetic energy is achieved. }
	\label{fig3}
\end{figure}

\begin{figure}[htbp]
	\centering
	\includegraphics[width=0.5\textwidth]{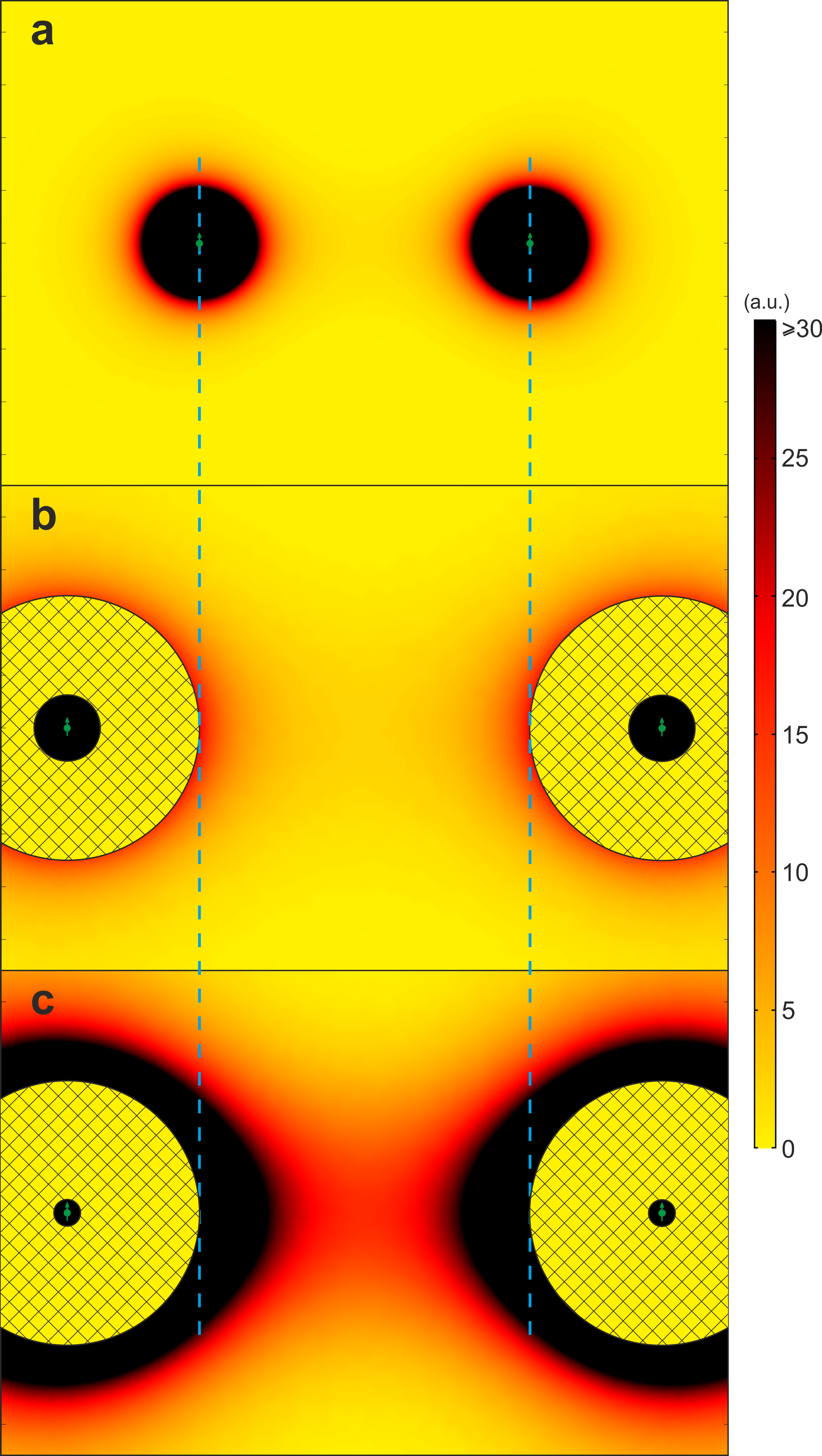}
	\caption{{\bf Enhanced magnetic coupling of two dipoles through free space.} In {\bf (a)}, magnetic energy density of two identical cylindrical dipoles separated a given gap. When separating and enclosing them with two of our shells with $R_2/R_1=4$ [{\bf (b)}], the magnetic energy density in the middle free space is similar to that in {\bf (a)}. When the inner radii of the shells are reduced to $R_2/R_1=10$ [{\bf (c)}], the magnetic energy is concentrated in the free space between the enclosed dipoles, enhancing the magnetic coupling.}
	\label{fig4}
\end{figure}

\begin{figure}[htbp]
	\centering
	\includegraphics[width=1.0\textwidth]{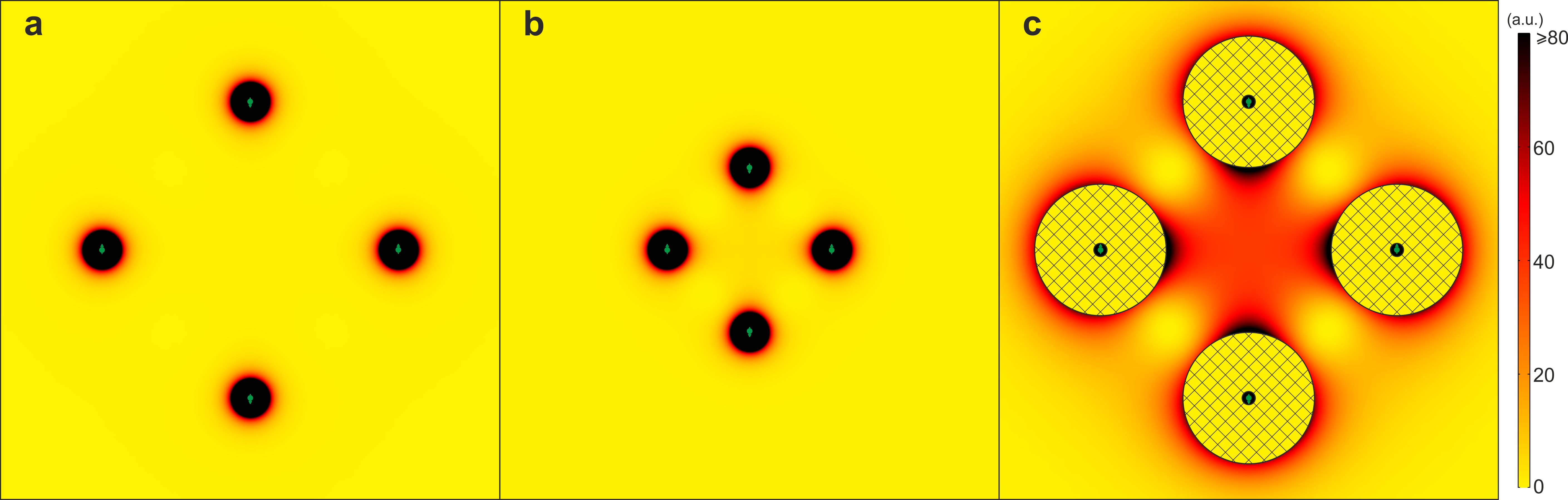}
	\caption{{\bf Magnetic field concentration in free space.} Four magnetic dipoles show a weak interaction when they 
are separated by free space (a) and the field in the region between them is very small (yellow colour), even when they are closer (b). However, when all four dipoles are surrounded 
by our magnetic shells [$R_2/R_1=10$] (c) the field in their exterior becomes greatly enhanced yielding a larger 
energy density in the intermediate region between the dipoles.}
	\label{fig5}
\end{figure}


\begin{thebibliography}{99}

\bibitem{TOpendry} Ward, A. J. and Pendry, J. B. Refraction and geometry in Maxwell�s equations. J. Mod. Opt. {\bf 43}, 773 (1996).

\bibitem{review_TO} Chen H., Chan C. T., and Sheng, P. Transformation optics and metamaterials. Nature Materials {\bf 9}, 387 (2010).

\bibitem{perfectlens} Pendry, J. B., Negative refraction makes a perfect lens. Phys. Rev. Lett. {\bf 18}, 3966 (2000).

\bibitem{controlling} Pendry, J. B., Schurig, D., and Smith, D. R. Controlling electromagnetic fields. Science 312, 1780 (2006). 
 
\bibitem{microwave} Schurig, D. et al. Metamaterial electromagnetic cloak at microwave frequencies.
Science 314 , 977 (2006).

\bibitem{wood} Wood B. and Pendry J. B. Metamaterials at zero frequency. J. Phys.: Condens. Matter {\bf 19}, 076208 (2007).


\bibitem{science} Gomory, F. et al. Experimental Realization of a Magnetic Cloak. Science, {\bf 335} 1466 (2012).

\bibitem{schuller} 
Schuller J. A., Barnard E. S., Cai W., Jun Y. C., White, J. S., and Brongersma, M. L. Plasmonics for extreme light concentration and manipulation.
Nature Materials {\bf 9}, 193 (2010). 

\bibitem{aubry} Aubry, A. et al., Plasmonic Light-Harvesting Devices over the Whole Visible Spectrum, Nano Letters {\bf 10}, 2574 (2010).


\bibitem{rahm} Rahm, M. et al. Design of electromagnetic cloaks and concentrators using form-invariant coordinate transformations of Maxwell's equations. Photon. Nanostruct.: Fundam. Applic. 6, 87 (2008).



\bibitem{plasmonics} Gramotnev, D. K. and  Bozhevolnyi, S. I. Plasmonics beyond the diffraction limit. 
Nature Photonics {\bf 4}, 83 (2010).


\bibitem{antimagnet} 
Sanchez A., Navau, C., Prat-Camps, J., and Chen, D.-X. Antimagnets: controlling magnetic fields with superconductor�metamaterial hybrids. New J. Phys. {\bf 13} 093034 (2011). 


\bibitem{lens} Choi S. et al. Magnetic lens effect using Gd-Ba-Cu-O bulk superconductor in very high magnetic field.
J. Appl. Phys. {\bf 111}, 07E728 (2012).

\bibitem{robbes} Robbes, D. Highly sensitive magnetometers - a Review. Sensors and Actuators A {\bf 129}, 86 (2006).


\bibitem{pannetier} Pannetier, M., Fermon, C., Le Goff, G., Simola, J., and Kerr, E. Femtotesla Magnetic Field Measurement with Magnetoresistive Sensors. Science {\bf 304}, 1648 (2004).



\bibitem{ripka} Ripka, P. and Janosek, M. Advances in Magnetic Field Sensors. IEEE Sensors Journal, {\bf 10}, 1108 (2010).


\bibitem{lenz2006} Lenz, J. and Edelstein, A. S. Magnetic Sensors and Their Applications. IEEE Sensors Journal {\bf 6}, 631 (2006).


\bibitem{griffith} 
Griffith, W. C., Jimenez-Martinez, R., Shah, V., Knappe, S., and
Kitching, J. Miniature atomic magnetometer integrated with flux concentrators.
Appl. Phys. Lett. {\bf 94}, 023502 (2009).



\bibitem{SQUIDreview} Kleiner, R., Koelle, D., Ludwig, L., and Clarke, J.
Superconducting Quantum Interference Devices:
State of the Art and Applications. Proceedings of the IEEE {\bf 92}, 1534 (2004).

\bibitem{clem} Babaei Brojeny, A. A., Mawatari, Y., Benkraouda M., and Clem, J. R. 
Magnetic fields and currents for two current-carrying parallel coplanar superconducting strips in a perpendicular magnetic field. Supercond. Sci. Technol. {\bf 15}, 1454 (2002).


\bibitem{biosensors} Wang S. X. and Li G. Advances in Giant Magnetoresistance Biosensors With Magnetic
Nanoparticle Tags: Review and Outlook. IEEE Trans. on Magn. {\bf 44}, 1687 (2008). 


\bibitem{pizzella} Pizzella, V., Della Penna, S., Del Gratta, C., and Romani, G. L. SQUID systems for biomagnetic imaging. Supercond. Sci. Technol. {\bf 14}, R79 (2001). 


\bibitem{fagaly} Fagaly, R. L. Superconducting quantum interference device instruments and applications. Rev. Sci. Instrum. {\bf 77}, 101101 (2006). 

\bibitem{microbead} Besse P.-A., Boero G., Demierre M., Pott V., and Popovic, R.
Detection of a single magnetic microbead using a miniaturized silicon Hall
sensor. Appl. Phys. Lett. {\bf 80}, 4199 (2002).

\bibitem{soljacic} Kurs, A. et al. Wireless Power Transfer via Strongly
Coupled Magnetic Resonances. Science {\bf 317}, 83 (2007).

\bibitem{smith} Huang, D. et al. Magnetic superlens-enhanced inductive coupling for wireless power transfer.
J. Appl. Phys. {\bf 111}, 064902 (2012).

\bibitem{wiltshire}
Wiltshire, M. C. K.  et al.
Microstructured Magnetic Materials for RF Flux Guides in Magnetic Resonance Imaging. Science {\bf 291}, 849 (2001).



\bibitem{trans} Kobayashi, M. and Pascual-Leone, A. Transcranial magnetic stimulation in neurology. Lancet Neurology {\bf 2}, 145 (2003).

\bibitem{bonmassar} Bonmassar, G. et al.
Microscopic magnetic stimulation of neural tissue. Nat. Comm. {\bf 3}, 921 (2012).


\end{thebibliography}
\end{document}